\documentclass[aps,pre,onecolumn,superscriptaddress,amsmath,amssymb]{revtex4-2}

\usepackage[utf8]{inputenc}
\usepackage{amsmath,amsthm}
\usepackage{graphicx}
\usepackage{easyReview}
\usepackage[colorlinks = true,
linkcolor = blue,
urlcolor = blue,
citecolor = blue,
anchorcolor = blue]{hyperref}
\usepackage{braket}
\usepackage{xcolor}
\global\long\def\al{\alpha}

\global\long\def\l{\lambda}

\global\long\def\ell#1{\theta_{#1}}
\global\long\def\bell#1{\tilde\theta_{#1}}

\def\ir{{\mathrm i}}
\def\eE{{\mathrm e}}
\def\no{\nonumber}

\renewcommand{\thetable}{\arabic{table}}

\theoremstyle{Plain}
\newtheorem*{rem}{Remark}
\newtheorem{hyp}{Hypothesis}

\begin{document}

\title{Hypotheses regarding Baxter's $T-Q$ relation for the periodic XYZ chain}

\author{Xin Zhang \footnote{Email:xinzhang@iphy.ac.cn}}
\affiliation{Beijing National Laboratory for Condensed Matter Physics, Institute of Physics, Chinese Academy of Sciences, Beijing 100190, China}

\begin{abstract}
Baxter's $T-Q$ relation for the periodic spin-$\frac12$ XYZ chain is studied. We extensively perform numerical calculations for the $T-Q$ relation and the Bethe ansatz equations. Numerical based hypotheses are then proposed to answer some open questions regarding Baxter's $T-Q$ relation and the XYZ chain.
\end{abstract}

\maketitle

\section{Introduction}

The spin-$\frac12$ XYZ chain is a typical integrable model in statistical mechanics, condensed matter physics and quantum information \cite{Baxter-book,Vsamaj—book}. It represents the most general form of the one-dimensional Heisenberg chain, where the exchange coefficients along the $x$, $y$ and $z$ directions are distinct. The XYZ model exhibits rich behavior due to the variation in exchange interactions along different axes. In the XYZ chain, the total magnetization $\sum_{j}\sigma_j^z$ is no longer conserved, implying significant difficulties in the study of this model.

Since Baxter's groundbreaking work on the integrability of the XYZ chain, numerous analytic approaches have been proposed for studying various aspects of the XYZ model, including the exact spectrum \cite{Baxter-book}, the eigenvector \cite{Baxter5,Baxter5a,Baxter6,Takhtadzhan1979,Deguchi02}, the ground state, and the excited states \cite{Klumper1988,Klumper1989}. The typical methods include the $T-Q$ relation \cite{Baxter-book,Fabricius07,FabMcCoy09,Baxter4,Yang2006}, the generalized coordinate Bethe ansatz \cite{Baxter5,Baxter5a,Baxter6,Zhang2024}, the generalized algebraic Bethe ansatz \cite{Takhtadzhan1979,Takhtajan1981,Fan1996,Slavnov2020}, the thermodynamic Bethe ansatz \cite{TBA,Takahashi1999,Xin2020}, and the off-diagonal Bethe ansatz \cite{Wang-book,Cao2013off,Cao2014}.

In this communication, we study the generic periodic spin-$\frac12$ XYZ chain with an even site number (denoted by 
$N$). The spectrum problem of this model was first solved by Baxter  \cite{Baxter-book}. Although the XYZ  does not have $U(1)$ symmetry, the eigenvalue of the quantum transfer matrix can still be parameterized by a homogeneous $T-Q$ relation where the number of Bethe roots is fixed to $N/2$. However, there are still some open questions. First, the Bethe vector of XYZ chain contains infinite series \cite{Takhtajan1981,Slavnov2020}, and its convergence is problematic. Moreover, there exist a parameter $\beta$ in the $T-Q$ relation and the corresponding Bethe vector, and deriving it value is always challenging. Another fact is that researchers often focus on the Hermitian case, and the corresponding result may not be applicable to the non-Hermitian scenario.

We extensively do numerical calculations across a range of XYZ models, exploring both the Hermitian and non-Hermitian cases. Based on the numerical results, we propose some hypotheses, including the value of the parameter $\beta$, the completeness of Baxter's $T-Q$ relation and the Bethe ansatz equations (BAEs), the existence of singular Bethe ansatz solution, the triangular limit of the XYZ model, the phantom string structure in the XXZ chain. The approach in our paper, i.e, proposing hypotheses based on convincing and enlightening numerical results, is common in theoretical physics, particularly when exact analytical solutions are difficult to obtain. Our results will be helpful to guide further theoretical and experimental investigations for the XYZ chain.

The paper is organized as follows. In Section \ref{XYZ}, we provide an introduction to the periodic spin-$\frac12$ XYZ chain and establish the notation used throughout the paper.
Section \ref{Integrability} is dedicated to demonstrating the integrability of the XYZ chain and presenting Baxter's $T-Q$ relation. In Section \ref{Hypotheses}, we put forth some hypotheses concerning Baxter's $T-Q$ relation. We delve into the XXZ limit (or trigonometric limit) and put forward another hypotheses in Section \ref{XXZ;Limit}.

\section{Periodic XYZ chain} \label{XYZ}
The Hamiltonian of the periodic spin-$\frac12$ XYZ chain is 
\begin{align} 
&H=\sum_{n=1}^{N}\left(J_x\,{\sigma}_n^x\sigma_{n+1}^x+J_y\,\sigma_n^y\sigma_{n+1}^y+J_z\,\sigma_n^z\sigma_{n+1}^z\right).\label{Hamiltonian}
\end{align}
Here, $N$ represents the length of the system and is assumed to be an even number, $\sigma^x,\sigma^y,\sigma^z$ are the Pauli matrices and the periodic boundary condition implies $\vec{\sigma}_{N+1}\equiv\vec\sigma_1$.  The exchange
coefficients $\{J_x,\,J_y,\,J_z\}$ are parameterized by the crossing parameter
$\eta$ \cite{Wang-book,OpenXYZ2022}
\begin{align}
J_x=\frac{\ell{4}(\eta)}{\ell{4}(0)},\quad J_y=\frac{\ell{3}(\eta)}{\ell{3}(0)},\quad J_z=\frac{\ell{2}(\eta)}{\ell{2}(0)},
\end{align}
where $\ell{\alpha}(u)\equiv\vartheta_{\al}(\pi
u,\eE^{\ir\pi\tau}),\,\alpha=1,2,3,4$ are elliptic theta functions and $\tau$ is a
complex number with a positive imaginary part. We adopt the notations of elliptic theta functions $\vartheta_{\al}(u,q)$ as introduced in Ref. \cite{WatsonBook}
\begin{align}
\begin{aligned}
&\vartheta_{1}(u,q)=2\sum_{n=0}^\infty(-1)^n q^{(n+\frac12)^2}\sin[(2n+1)u],\\
&\vartheta_{2}(u,q)=2\sum_{n=0}^\infty q^{(n+\frac12)^2}\cos[(2n+1)u],\\
&\vartheta_{3}(u,q)=1+2\sum_{n=1}^\infty q^{n^2}\cos(2nu),\\
&\vartheta_{4}(u,q)=1+2\sum_{n=1}^\infty (-1)^nq^{n^2}\cos(2nu).
\end{aligned}
\end{align}
In this paper, we will consider a generic XYZ model, where $\eta$ is not a root of unity, as follows 
\begin{align}
\eta\neq \frac{2K+2L\tau}{P},\quad P,K,L\in\mathbb{Z}.
\end{align}
\begin{rem}
When $\tau$ is purely imaginary and $\eta$ is either real or purely imaginary, the Hamiltonian in (\ref{Hamiltonian}) is Hermitian. In the following text, we will consider some non-Hermitian cases with generic complex $\tau$ and $\eta$ to validate the accuracy of our numerical results. 
\end{rem}

\section{Integrability \& Baxter's $T-Q$ relation}\label{Integrability}

The $R$-matrix of the XYZ model is \cite{Baxter-book}
\begin{align}
R(u)=\begin{pmatrix}
\alpha_1(u) & 0 & 0 & \alpha_4(u) \\
0 & \alpha_2(u) & \alpha_3(u) & 0\\
0 & \alpha_3(u) & \alpha_2(u) & 0\\
\alpha_4(u) & 0 & 0 & \alpha_1(u)
\end{pmatrix},\label{R;XYZ}
\end{align}
where $u$ is the spectral parameter and the non-zero entries in $R(u)$ are 
\begin{align}
\alpha_1(u)=\frac{\bell{4}(u)\bell{1}(u+\eta)}{\bell{4}(0)\bell{1}(\eta)},\qquad
\alpha_2(u)=\frac{\bell{1}(u)\bell{4}(u+\eta)}{\bell{4}(0)\bell{1}(\eta)},\\
\alpha_3(u)=\frac{\bell{4}(u)\bell{4}(u+\eta)}{\bell{4}(0)\bell{4}(\eta)},\qquad
\alpha_4(u)=\frac{\bell{1}(u)\bell{1}(u+\eta)}{\bell{4}(0)\bell{4}(\eta)}.
\end{align}
Here, $\bell{\alpha}(u)\equiv\vartheta_{\al}(\pi
u,\eE^{2\ir\pi\tau}),\,\alpha=1,2,3,4$.

The $R$-matrix in Eq. (\ref{R;XYZ}) satisfies the Yang-Baxter equation (YBE)
\begin{align}
R_{1,2}(u_1-u_2)R_{1,3}(u_1-u_3)R_{2,3}(u_2-u_3)=R_{2,3}(u_2-u_3)R_{1,3}(u_1-u_3)R_{1,2}(u_1-u_2).
\end{align}  
Introduce the quantum transfer matrix of the periodic XYZ chain
\begin{align}
t(u)=tr_0\{T_0(u)\},\qquad T_0(u)=R_{0,N}(u)\cdots R_{0,1}(u).\label{TransferMatrix}
\end{align}
The Hamiltonian is obtained in terms of the transfer matrix $t(u)$ as 
\begin{align}
H=2\,\frac{\ell{1}(\eta)}{\ell{1}'(0)}\left.\frac{\partial\ln t(u)}{\partial u}\right|_{u=0}-N\frac{\ell{1}'(\eta)}{\ell{1}'(0)}.
\end{align}

The eigenvalue of $t(u)$, denoted by $\Lambda(u)$, can be parameterized by Baxter's homogeneous $T-Q$ relation \cite{Baxter-book}
\begin{align}
\Lambda(u)&=\eE^{\beta\gamma}\,\frac{\ell{1}^N(u+\eta)}{\ell{1}^N(\eta)}\frac{Q(u-\eta)}{Q(u)}+\eE^{-\beta\gamma}\,\frac{\ell{1}^N(u)}{\ell{1}^N(\eta)}\frac{Q(u+\eta)}{Q(u)},\quad\gamma=\ir\pi\eta,\label{TQ1}\\
Q(u)&=\prod_{j=1}^{M}\ell{1}(u-\l_j+\tfrac{\eta}{2}),\qquad M=\tfrac{N}{2}.
\end{align}
To ensure that $\Lambda(u)$ is an entire function of $u$, we get the following associated BAEs
\begin{align}
&\eE^{2\beta\gamma}\,\ell{1}^N(\l_j+\tfrac{\eta}{2})\prod_{k\neq j}^{M}\ell{1}(\l_j-\l_k-\eta)=\ell{1}^N(\l_j-\tfrac{\eta}{2})\prod_{k\neq j}^{M}\ell{1}(\l_j-\l_k+\eta),\label{BAE}\\
&\mbox{or}\,\,\,\eE^{2\beta\gamma}\,\left[\frac{\ell{1}(\l_j+\frac{\eta}{2})}{\ell{1}(\l_j-\frac{\eta}{2})}\right]^N\prod_{k\neq j}^{M}\frac{\ell{1}(\l_j-\l_k-\eta)}{\ell{1}(\l_j-\l_k+\eta)}=1,\quad j=1,2,\ldots,M,\quad M=\tfrac{N}{2}.\label{BAE;M;1}
\end{align}
The valid Bethe
roots should satisfy an additional sum rule \cite{Baxter-book,Baxter4}
\begin{align}
2\sum_{j=1}^{M}\l_j=k+p\tau,\quad k,p\in\mathbb{Z}.\label{SelectionRule}
\end{align} 
The energy of the system in terms of Bethe roots is 
\begin{align}
&E(\l_1,\ldots,\l_{M})=2\sum_{j=1}^{M}[g(\l_j-\tfrac{\eta}{2})-g(\l_j+\tfrac{\eta}{2})]+Ng(\eta),\quad g(u)=\frac{\ell{1}(\eta)\ell{1}'(u)}{\ell{1}'(0)\ell{1}(u)}.\label{Energy}
\end{align}

The Bethe vector of the XYZ model can be constructed as an infinite series using the generalized algebraic Bethe ansatz method \cite{Takhtadzhan1979,Takhtajan1981}
\begin{align}
\ket{\Psi_\beta(\l_1,\ldots,\l_M)}=\sum_{l=-\infty}^\infty \eE^{-l\gamma\beta}\,\mathcal{B}_{l-1,l+1}(\l_1-\tfrac{\eta}{2})\cdots \mathcal{B}_{l-M,l+M}(\l_M-\tfrac{\eta}{2})\ket{\Omega^{l-M}},\label{BetheVector}
\end{align} 
where $\mathcal{B}_{l-1,l+1}(\l_1-\tfrac{\eta}{2})$ is a non-diagonal element of the gauge-transformed monodromy matrix and $\ket{\Omega^{l}}$ represents a $l$-dependent factorized reference vector. More details regarding the construction of the Bethe vector can be located in Refs. \cite{Slavnov2020,kulkarni2023}.

We solve the BAEs for various system parameters numerically. Once the Bethe roots are obtained, they can be substituted into the $T-Q$ relation (\ref{TQ1}) and the energy formula (\ref{Energy}). We compare the results with those obtained from exact diagonalization techniques. This helps us validate our numerical solutions of the BAEs. Based on the convincing and enlightening numerical results, we present several interesting hypotheses in Sections \ref{Hypotheses} and \ref{XXZ;Limit}..
                                          \begin{rem}
It can be verified that $\{\l_1,\ldots,\l_M,\beta\}$, $\{\l_1+1,\ldots,\l_M,\beta\}$ and $\{\l_1+\tau,\ldots,\l_M,\beta+2\}$  represent equivalent solutions. When $\l_j\to\l_j+\tau$, it can be proven that $\beta\to\beta+2$, $p\to p+2$. Therefore, there exists a direct correspondence between the parameters $\beta$ and $p$.
\end{rem}                                                                        \section{Hypotheses for Baxter's $T-Q$ relation} \label{Hypotheses}         

\begin{hyp}\label{hyp1}
For generic periodic XYZ chain, the parameter $\beta$ in BAE (\ref{BAE;M;1}) is equal to the integer $p$ in the sum rule (\ref{SelectionRule}). Therefore, we can rewrite the sum rule (\ref{SelectionRule}) as 
\begin{align}
\sin\left(\left(2\sum_{j=1}^M\l_j-\beta\tau\right)\pi\right)=0.\label{SumRule}
\end{align}
\end{hyp}
The numerical results of small-size systems presented in Tables \ref{Tab1}, \ref{Tab1-1}, \ref{Tab2}, \ref{4.a}, and \ref{4.b} provide compelling evidence  supporting the aforementioned hypothesis. Another argument is the analytic result for $N=2$ case, where the solutions of BAE are \cite{Slavnov2020}
\begin{align}
&\l_1=0,\frac{1}{2},\quad \beta=0, \no\\
&\l_1=\frac{\tau}{2},\frac{1}{2}+\frac{\tau}{2},\quad \beta=1.
\end{align}

The hypothesis in \ref{hyp1} also implies that the Bethe vector in Eq.  (\ref{BetheVector}) should converge when the Bethe roots $\{\l_1,\ldots,\l_M\}$ satisfy the BAEs and $\beta$ equals $p$. 

\begin{rem}
The $M+1$ unknowns $\{\l_1,\ldots,\l_M,\beta\}$ are now completely determined by a set of $M+1$ equations, encompassing both the BAEs (\ref{BAE}) and the additional sum rule (\ref{SumRule}). 
\end{rem}

\begin{table}[htbp]
\begin{minipage}{0.45\textwidth}
\begin{tabular}{|c|c|c|r|}
\hline 
$\l_1$ & $\l_2$ & $\beta$ & $E$~~~~ \\
\hline 
0.0877$\ir$ & $-$0.0877$\ir$ & 0 & $-$7.8613 \\
0 & $\frac{\tau}{2}$ & 1 & $-$6.4147 \\
0 & $\frac12$+$\frac{\tau}{2}$ & 1 & $-$2.6983 \\
0 & $\frac12$ & 0 & $-$2.0665 \\
$-$0.1093$\ir$ & $\frac12$+0.1093$\ir$ & 0 & 0.0000 \\
0.1093$\ir$ & $\frac12-$0.1093$\ir$ & 0 & 0.0000 \\
0.1180$\ir$ & $\frac12$+0.1820$\ir$ & 1 & 0.0000 \\
$-$0.1180$\ir$ & $\frac12-$0.1820$\ir$ & $-$1 & 0.0000 \\
0.8505$-\frac{\tau}{4}$ & 0.1495$-\frac{\tau}{4}$ & $-$1 & 0.0000 \\
0.8505+$\frac{\tau}{4}$ & 0.1495+$\frac{\tau}{4}$ & 1 & 0.0000 \\
$\frac{\eta}{2}$ & $-\frac{\eta}{2}$ & 0 & 0.0000 \\
0.8229+$\frac{\tau}{2}$ & 0.1771+$\frac{\tau}{2}$ & 2 & 1.4107 \\
$\frac{\tau}{2}$ & $\frac12$+$\frac{\tau}{2}$ & 2 & 2.0665 \\
$\frac12$ & $\frac{\tau}{2}$ & 1 & 2.6983 \\
$\frac12$ & $\frac12$+$\frac{\tau}{2}$ & 1 & 6.4147 \\
$\frac12-$0.1391$\ir$ & $\frac12$+0.1391$\ir$ & 0 & 6.4506 \\
\hline 
\end{tabular}
\end{minipage}
\begin{minipage}{0.45\textwidth}
\begin{tabular}{|c|c|c|r|}
\hline 
$\l_1$ & $\l_2$ & $\beta$ & $E$~~~~ \\
\hline 
0.0876 & 0.9124 & 0 & $-$9.2437 \\
0 & $\frac12$ & 0 & $-$6.8499 \\
0 & $\frac12$+$\frac{\tau}{2}$ & 1 & $-$6.5659 \\
$\frac12$ & $\frac12$+$\frac{\tau}{2}$ & 1 & $-$0.4967 \\
$\frac12$$-$0.1503$\ir$ & $\frac12$+0.1503$\ir$ & 0 & $-$0.4962 \\
0.7500+0.1751$\ir$ & 0.7500$-$0.1751$\ir$ & 0 & 0.0000 \\
0.2500+0.1751$\ir$ & 0.2500$-$0.1751$\ir$ & 0 & 0.0000 \\
0.2735 & 0.2265+$\frac{\tau}{2}$ & 1 & 0.0000 \\
0.7265 & 0.7735+$\frac{\tau}{2}$ & 1 & 0.0000 \\
0.0840 & 0.9160+$\frac{\tau}{2}$ & 1 & 0.0000 \\
0.9160 & 0.0840+$\frac{\tau}{2}$ & 1 & 0.0000 \\
$\frac{\eta}{2}$ & $-$$\frac{\eta}{2}$ & 0 & 0.0000 \\
0 & $\frac{\tau}{2}$ & 1 & 0.4967 \\
$\frac12$ & $\frac{\tau}{2}$ & 1 & 6.5659 \\
$\frac{\tau}{2}$ & $\frac12$+$\frac{\tau}{2}$ & 2 & 6.8499 \\
0.9194+$\frac{\tau}{2}$ & 0.0806+$\frac{\tau}{2}$ & 2 & 9.7400 \\
\hline 
\end{tabular}
\end{minipage}
\caption{Left: Numerical solutions of BAEs
(\ref{BAE;M;1}) with $N=4,\,\tau=0.6\ir$, $\eta=\frac{\pi}{10}$. Right: Numerical solutions of BAEs (\ref{BAE;M;1}) with $N=4$, $\tau=0.6\ir$ $\eta=\frac{\ir\pi}{10}$.}
\label{Tab1}
\end{table}

\begin{table}
\begin{tabular}{|c|c|c|c|}
\hline 
$\l_1$ & $\l_2$ & $\beta$ & $E$ \\
\hline
 0.9081+0.0942$\ir$ & 0.0919$-$0.0942$\ir$ & 0 & $-$5.4687+4.7280$\ir$ \\
 0 & $\frac{1}{2}+\frac{\tau}{2}$ & 1 & $-$4.4309+4.6335$\ir$ \\
 0 & $\frac12$ & 0 & $-$2.9677+5.0873$\ir$ \\
 0 & $\frac{\tau}{2}$ & 1 & $-$0.7807$-$4.5312$\ir$ \\
 0.4822$-$0.3074$\ir$ & 0.0178+0.3074$\ir$ & 0 & 0.0000 \\
 0.5178+0.3074$\ir$ & 0.9822$-$0.3074$\ir$ & 0 & 0.0000 \\
 0.3003+0.2579$\ir$ & 0.8997+0.0421$\ir$ & 1 & 0.0000 \\
 0.6997$-$0.2579$\ir$ & 0.1003$-$0.0421$\ir$ & $-$1 & 0.0000 \\
 0.6382+0.0369$\ir$ & 0.0618+0.2631$\ir$ & 1 & 0.0000 \\
 0.3618$-$0.0369$\ir$ & 0.9382$-$0.2631$\ir$ & $-$1 & 0.0000 \\
 $\frac{\eta}{2}$ & $-\frac{\eta}{2}$ & 0 & 0.0000 \\
 $\frac12$ & $\frac{1}{2}+\frac{\tau}{2}$ & 1 & 0.7807+4.5312$\ir$ \\
 0.5885+0.1360$\ir$ & 0.4115$-$0.1360$\ir$ & 0 & 0.8989+4.5545$\ir$ \\
 $\frac{\tau}{2}$ & $\frac{1}{2}+\frac{\tau}{2}$ & 2 & 2.9677$-$5.0873$\ir$ \\
 $\frac12$ & $\frac{\tau}{2}$ & 1 & 4.4309$-$4.6335$\ir$ \\
 0.8829$-$0.3103$\ir$ & 0.1171+0.3103$\ir$ & 0 & 4.5698$-$9.2825$\ir$ \\
\hline
\end{tabular}
\caption{Numerical solutions of BAEs
 (\ref{BAE;M;1}) with $N=4,\,\tau=0.4+0.6\ir$, $\eta=\frac{1}{\eE}+\frac{\ir\pi}{10}$.}
\label{Tab1-1}
\end{table}

\begin{table}[htbp]
\begin{tabular}{|c|c|c|c|r|}
\hline 
$\l_1$ & $\l_2$ & $\l_3$ & $\beta$ &$E$~~~~ \\
\hline 
0 & $-$0.0991$\ir$ & 0.0991$\ir$ & 0 & $-$10.3014 \\
$\frac{\tau}{2}$ & $-$0.0383$\ir$ & 0.0383$\ir$ &  1 & $-$8.0739 \\
$\frac12$+$\frac{\tau}{2}$ & 0.0382$\ir$ & $-$0.0382$\ir$ & 1 & $-$8.0306 \\
$\frac12$ & 0.0357$\ir$ & $-$0.0357$\ir$ &  0 & $-$7.2138 \\
 $\frac{\eta}{2}$ & $-$$\frac{\eta}{2}$ & 0 & 0 & $-$5.4956 \\
0.0175$\ir$ & 0.1183$\ir$ & $\frac12$$-$0.1358$\ir$ & 0 & $-$4.7721 \\
$-$0.0175$\ir$ & $-$0.1183$\ir$ & $\frac12$+0.1358$\ir$ & 0 & $-$4.7721 \\
0.1462$\ir$ & $\frac12$+0.7299$\ir$ & 0.0239$\ir$ & 1 & $-$4.6690 \\
$-$0.1462$\ir$ & $\frac12$$-$0.7299$\ir$ & $-$0.0239$\ir$ & $-$1 & $-$4.6690 \\
$-$0.7263$\ir$ & $-$0.1495$\ir$ & $-$0.0242$\ir$ & $-$1 & $-$4.6653 \\
0.7263$\ir$ & 0.1495$\ir$ & 0.0242$\ir$ & 1 & $-$4.6653 \\
$-$0.0453$\ir$ & 0.1593$\ir$ & 0.7859$\ir$ & 1 & $-$3.7781 \\
0.0453$\ir$ & $-$0.1593$\ir$ & $-$0.7859$\ir$ & $-$1 & $-$3.7781 \\
0.1568$\ir$ & $\frac12$+0.7886$\ir$ & $-$0.0453$\ir$ & 1 & $-$3.7684 \\
$-$0.1568$\ir$ & $\frac12$$-$0.7886$\ir$ & 0.0453$\ir$ & $-$1 & $-$3.7684 \\
0.0464$\ir$ & $-$0.1291$\ir$ & $\frac12$+0.0827$\ir$ & 0 & $-$3.4956 \\
$-$0.0464$\ir$ & 0.1291$\ir$ & $\frac12$$-$0.0827$\ir$ & 0 & $-$3.4956 \\
0 & 0.2462+$\frac{\tau}{2}$ & 0.7538+$\frac{\tau}{2}$ & 2 & $-$2.5054 \\
0 & $\frac{\tau}{2}$ & $\frac12$+$\frac{\tau}{2}$ &  2 & $-$2.5048 \\
0.1085$\ir$ & 0.8844$-$0.0542$\ir$ & 0.1156$-$0.0542$\ir$ & 0 & $-$1.8716 \\
$-$0.1085$\ir$ & 0.8844+0.0542$\ir$ & 0.1156+0.0542$\ir$ & 0 & $-$1.8716 \\
 0 & $\frac12$ & $\frac{\tau}{2}$ & 1 & $-$1.5201 \\
0 & $\frac12$ & $\frac12$+$\frac{\tau}{2}$ & 1 & $-$1.4707 \\
0 & $\frac12$$-$0.2036$\ir$ & $\frac12$+0.2036$\ir$ & 0 & $-$1.1901 \\
$-$0.0710$\ir$ & 0.2464$-$0.8645$\ir$ & 0.7536$-$0.8645$\ir$ & $-$2 & $-$0.5051 \\
0.0710$\ir$ & 0.2464+0.8645$\ir$ & 0.7536+0.8645$\ir$ & 2 & $-$0.5051 \\
$-$0.8647$\ir$ & $\frac12$$-$0.8643$\ir$ & $-$0.0710$\ir$ & $-$2 & $-$0.5048 \\
0.8647$\ir$ & $\frac12$+0.8643$\ir$ & 0.0710$\ir$ & 2 & $-$0.5048 \\
$-$0.0638$\ir$ & $\frac12$+0.0375$\ir$ & $-$0.8738$\ir$ & $-$1 & 0.1682 \\
0.0638$\ir$ & $\frac12$$-$0.0375$\ir$ & 0.8738$\ir$ & 1 & 0.1682 \\
0.0636$\ir$ & $\frac12$+0.8737$\ir$ &  $\frac12$$-$0.0372$\ir$ & 1 & 0.2108 \\
$-$0.0636$\ir$ & $\frac12$$-$0.8737$\ir$ & $\frac12$+0.0372$\ir$ & $-$1 & 0.2108 \\
\hline
\end{tabular}
\begin{tabular}{|c|c|c|c|c|}
\hline 
$\l_1$ & $\l_2$ & $\l_3$ & $\beta$ &$E$ \\
\hline
$\frac12$ & $-$0.1436$\ir$ & 0.1436$\ir$ & 0 & 0.3661 \\
$\frac12$+$\frac{\tau}{2}$ & 0.1678$\ir$ & $-$0.1678$\ir$ & 1 & 0.4154 \\
$\frac{\tau}{2}$ & 0.1691$\ir$ & $-$0.1691$\ir$ &  1 & 0.4170 \\
0.0628$\ir$ & $\frac12$$-$0.2291$\ir$ & $\frac12$+0.1663$\ir$ &  0 & 0.5051 \\
$-$0.0628$\ir$ & $\frac12$+0.2291$\ir$ & $\frac12$$-$0.1663$\ir$ &  0 & 0.5051 \\
$\frac{\eta}{2}$ & $-$$\frac{\eta}{2}$ & $\frac{\tau}{2}$ & 1 & 1.4707 \\
$\frac{\eta}{2}$ & $-$$\frac{\eta}{2}$ & $\frac{1}{2}+\frac{\tau}{2}$ & 1 & 1.5201 \\
$\frac{\eta}{2}$ & $-\frac{\eta}{2}$ & $\frac{1}{2}$ & 0 & 2.5048 \\
$\frac12$$-$0.1907$\ir$ & 0.1162+0.0954$\ir$ & 0.8838+0.0954$\ir$ &  0 & 2.7721 \\
$\frac12$+0.1907$\ir$ & 0.1162$-$0.0954$\ir$ & 0.8838$-$0.0954$\ir$ &  0 & 2.7721 \\
$\frac12$+0.5844$\ir$ & 0.8773+0.1578$\ir$ & 0.1227+0.1578$\ir$ &  1 & 2.9806 \\
$\frac12$$-$0.5844$\ir$ & 0.8773$-$0.1578$\ir$ & 0.1227$-$0.1578$\ir$ &  $-$1 & 2.9806 \\
$-$0.5624$\ir$ & 0.8734$-$0.1688$\ir$ & 0.1266$-$0.1688$\ir$ &  $-$1 & 2.9837 \\
0.5624$\ir$ & 0.8734+0.1688$\ir$ & 0.1266+0.1688$\ir$ &  1 & 2.9837 \\
0 & 0.7672 & 0.2328 & 0 & 3.0057 \\
0.2512$\ir$ & 0.7499+0.7744$\ir$ & 0.2501+0.7744$\ir$ &  2 & 3.4955 \\
$-$0.2512$\ir$ & 0.7499$-$0.7744$\ir$ & 0.2501$-$0.7744$\ir$ &  $-$2 & 3.4955 \\
$-$0.2514$\ir$ & $\frac12$$-$0.7765$\ir$ & $-$0.7721$\ir$ & $-$2 & 3.4956 \\
0.2514$\ir$ & $\frac12$+0.7765$\ir$ & 0.7721$\ir$ &  2 & 3.4956 \\
$-$0.7855$\ir$ & $\frac12$+0.0786$\ir$ & $-$0.1931$\ir$ & $-$1 & 3.7684 \\
0.7855$\ir$ & $\frac12$$-$0.0786$\ir$ & 0.1931$\ir$ & 1 & 3.7684 \\
0.1895$\ir$ & $\frac12$+0.7889$\ir$ & $\frac12$$-$0.0784$\ir$ &  1 & 3.7781 \\
$-$0.1895$\ir$ & $\frac12$$-$0.7889$\ir$ & $\frac12$+0.0784$\ir$ &  $-$1 & 3.7781 \\
$-$0.1672$\ir$ & $\frac12$$-$0.0910$\ir$ &  $\frac12$+0.2582$\ir$ & 0 & 3.8717 \\
0.1672$\ir$ & $\frac12$+0.0910$\ir$ &  $\frac12$$-$0.2582$\ir$ & 0 & 3.8717 \\
$\frac{\tau}{2}$ & 0.6683+$\frac{\tau}{2}$ & 0.3317+$\frac{\tau}{2}$ & 3 & 4.4862 \\
$\frac12$+$\frac{\tau}{2}$ & 0.8349+$\frac{\tau}{2}$ & 0.1651+$\frac{\tau}{2}$ & 3 & 4.4862 \\
$\frac12$ & 0.7562+$\frac{\tau}{2}$ & 0.2438+$\frac{\tau}{2}$ &  2 & 5.4942 \\
$\frac12$ & $\frac{\tau}{2}$ & $\frac12$+$\frac{\tau}{2}$ &  2 & 5.4956 \\
$\frac{\tau}{2}$ & $\frac12$+0.1363$\ir$ & $\frac12$$-$0.1363$\ir$ &  1 & 6.1122 \\
$\frac12$+$\frac{\tau}{2}$ & $\frac12$+0.1347$\ir$ & $\frac12$$-$0.1347$\ir$ &  1 & 6.1692 \\
$\frac12$ & $\frac12$$-$0.2977$\ir$ & $\frac12$+0.2977$\ir$ & 0 & 6.3631 \\
\hline
\end{tabular}
\caption{Numerical solutions of BAEs (\ref{BAE;M;1})  with $N=6,\,\tau=1.8\ir,\,\eta=\frac{\pi}{5 \eE}$.}\label{Tab2}
\end{table}

\begin{table}[htbp]
\renewcommand{\thetable}{4.a}
\begin{tabular}{|c|c|c|c|c|}
\hline 
$\l_1$ & $\l_2$ & $\l_3$ & $\beta$ &$E$ \\
\hline 
 0 & 0.1325$-$0.1429$\ir$ & 0.8675+0.1429$\ir$ & 0 & $-$7.7107+6.8714$\ir$ \\
 $\frac{1}{2}+\frac{\tau}{2}$ & 0.0546$-$0.0648$\ir$ & 0.9454+0.0648$\ir$ & 1 & $-$7.2495+6.8431$\ir$ \\
 0.0581$-$0.0727$\ir$ & 0.9419+0.0727$\ir$ & $\frac12$ & 0 & $-$5.5796+7.1659$\ir$ \\
 $\frac{\eta}{2}$ & $-\frac{\eta}{2}$ & 0 & 0 & $-$4.0896+2.5948$\ir$ \\
 0.1418$-$0.1307$\ir$ & 0.6461$-$0.2161$\ir$ & 0.0121+0.0468$\ir$ & $-$1 & $-$3.3265+4.4067$\ir$ \\
 0.8582+0.1307$\ir$ & 0.3539+0.2161$\ir$ & 0.9879$-$0.0468$\ir$ & 1 & $-$3.3265+4.4067$\ir$ \\
 0 & $\frac12$ & $\frac{1}{2}+\frac{\tau}{2}$ & 1 & $-$3.3089+7.1260$\ir$ \\
 0 & 0.5765+0.1293$\ir$ & 0.4235$-$0.1293$\ir$ & 0 & $-$3.0088+7.1181$\ir$ \\
 0.4203+0.2320$\ir$ & 0.9328+0.0421$\ir$ & 0.5469+0.3259$\ir$ & 2 & $-$2.9472+4.5576$\ir$ \\
 0.5797$-$0.2320$\ir$ & 0.0672$-$0.0421$\ir$ & 0.4531$-$0.3259$\ir$ & $-$2 & $-$2.9472+4.5576$\ir$ \\
 0.0469+0.2719$\ir$ & 0.0059+0.0835$\ir$ & 0.6473$-$0.0554$\ir$ & 1 & $-$2.8377+2.7624$\ir$ \\
 0.9531$-$0.2719$\ir$ & 0.9941$-$0.0835$\ir$ & 0.3527+0.0554$\ir$ & $-$1 & $-$2.8377+2.7624$\ir$ \\
 0.6154+0.0470$\ir$ & 0.0695$-$0.0430$\ir$ & 0.0151+0.2960$\ir$ & 1 & $-$2.6441+4.4197$\ir$ \\
 0.3846$-$0.0470$\ir$ & 0.9305+0.0430$\ir$ & 0.9849$-$0.2960$\ir$ & $-$1 & $-$2.6441+4.4197$\ir$ \\
 0.9427$-$0.0919$\ir$ & 0.2411$-$0.1397$\ir$ & 0.3162+0.2316$\ir$ & 0 & $-$2.5808+2.9434$\ir$ \\
 0.0573+0.0919$\ir$ & 0.7589+0.1397$\ir$ & 0.6838$-$0.2316$\ir$ & 0 & $-$2.5808+2.9434$\ir$ \\
 0.1366$-$0.1023$\ir$ & 0.7274$-$0.2791$\ir$ & 0.9361+0.0814$\ir$ & $-$1 & $-$2.3161+0.1055$\ir$ \\
 0.8634+0.1023$\ir$ & 0.2726+0.2791$\ir$ & 0.0639$-$0.0814$\ir$ & 1 & $-$2.3161+0.1055$\ir$ \\
 $\frac{\tau}{2}$ & 0.1077$-$0.0741$\ir$ & 0.8923+0.0741$\ir$ & 1 & $-$2.1256$-$1.5823$\ir$ \\
 0.1656+0.0067$\ir$ & 0.8344$-$0.0067$\ir$ & $\frac{\tau}{2}$ & 1 & $-$1.9583+2.7888$\ir$ \\
 0.0336$-$0.1977$\ir$ & 0.4020$-$0.0552$\ir$ & 0.5644+0.2529$\ir$ & 0 & $-$1.4460+4.6528$\ir$ \\
 0.9664+0.1977$\ir$ & 0.5980+0.0552$\ir$ & 0.4356$-$0.2529$\ir$ & 0 & $-$1.4460+4.6528$\ir$ \\
 0.0171$-$0.1951$\ir$ & 0.4708+0.0810$\ir$ & 0.3121$-$0.1858$\ir$ & $-$1 & $-$1.4378+4.6199$\ir$ \\
 0.9829+0.1951$\ir$ & 0.5292$-$0.0810$\ir$ & 0.6879+0.1858$\ir$ & 1 & $-$1.4378+4.6199$\ir$ \\
 0.1034+0.2082$\ir$ & 0.7361$-$0.1064$\ir$ & 0.1605$-$0.1018$\ir$ & 0 & $-$1.3950+3.4037$\ir$ \\
 0.8966$-$0.2082$\ir$ & 0.2639+0.1064$\ir$ & 0.8395+0.1018$\ir$ & 0 & $-$1.3950+3.4037$\ir$ \\
 0.0530+0.0037$\ir$ & 0.7231$-$0.2704$\ir$ & 0.7239+0.2667$\ir$ & 0 & $-$1.2051+0.5145$\ir$ \\
 0.9470$-$0.0037$\ir$ & 0.2769+0.2704$\ir$ & 0.2761$-$0.2667$\ir$ & 0 & $-$1.2051+0.5145$\ir$ \\
 0 & $\frac{\tau}{2}$ & $\frac{1}{2}+\frac{\tau}{2}$ & 2 & $-$1.1220$-$2.4925$\ir$ \\
 $\frac{\eta}{2}$ & $-\frac{\eta}{2}$ & $\frac{1}{2}+\frac{\tau}{2}$ & 1 & $-$0.3412+2.0387$\ir$ \\
 0.5969$-$0.3263$\ir$ & 0.4031+0.3263$\ir$ & $\frac12$ & 0 & $-$0.1009+0.8699$\ir$ \\
 $\frac{1}{2}+\frac{\tau}{2}$ & 0.0050$-$0.3047$\ir$ & 0.5950$-$0.2953$\ir$ & $-$1 & $-$0.0627+0.8090$\ir$ \\
\hline
\end{tabular}
\caption{Numerical solutions of BAEs (\ref{BAE;M;1})  with $N=6,\,\tau=0.4+0.6\ir,\,\eta=\frac{1}{\eE}+\frac{\ir\pi}{10}$.}\label{4.a}
\end{table}
\begin{table}[htbp]
\renewcommand{\thetable}{4.b}
\begin{tabular}{|c|c|c|c|c|}
\hline 
$\l_1$ & $\l_2$ & $\l_3$ & $\beta$ &$E$ \\
\hline 
 0.9564$-$0.3600$\ir$ & 0.0436+0.3600$\ir$ & 0 & 0 & 0.2387+0.7843$\ir$ \\
 0 & $\frac12$ & $\frac{\tau}{2}$  & 1 & 0.3412$-$2.0387$\ir$ \\
 0.0640+0.0013$\ir$ & 0.7205$-$0.2688$\ir$ & 0.5154$-$0.0325$\ir$ & $-$1 & 0.3954+1.1998$\ir$ \\
 0.9360$-$0.0013$\ir$ & 0.2795+0.2688$\ir$ & 0.4846+0.0325$\ir$ & 1 & 0.3954+1.1998$\ir$ \\
 $\frac{\eta}{2}$ & $-\frac{\eta}{2}$ & $\frac12$ & 0 & 1.1220+2.4925$\ir$ \\
 $\frac{1}{2}+\frac{\tau}{2}$ & 0.5518+0.0922$\ir$ & 0.4482$-$0.0922$\ir$ & 1 & 1.2769+6.7410$\ir$ \\
 0.3832$-$0.1908$\ir$ & 0.6168+0.1908$\ir$ & $\frac12$ & 0 & 1.2927+6.7368$\ir$ \\
 0.0786$-$0.0401$\ir$ & 0.2586+0.2953$\ir$ & 0.6628$-$0.2552$\ir$ & 0 & 1.4460$-$4.6528$\ir$ \\
 0.9214+0.0401$\ir$ & 0.7414$-$0.2953$\ir$ & 0.3372+0.2552$\ir$ & 0 & 1.4460$-$4.6528$\ir$ \\
 0.8848$-$0.1680$\ir$ & 0.6626$-$0.2779$\ir$ & 0.2526+0.1459$\ir$ & $-$1 & 1.4554$-$1.9007$\ir$ \\
 0.1152+0.1680$\ir$ & 0.3374+0.2779$\ir$ & 0.7474$-$0.1459$\ir$ & 1 & 1.4554$-$1.9007$\ir$ \\
 0.1081+0.2954$\ir$ & 0.3500$-$0.2926$\ir$ & 0.5419$-$0.0028$\ir$ & 0 & 2.2504$-$0.0822$\ir$ \\
 0.8919$-$0.2954$\ir$ & 0.6500+0.2926$\ir$ & 0.4581+0.0028$\ir$ & 0 & 2.2504$-$0.0822$\ir$ \\
 0 & 0.7545$-$0.2870$\ir$ & 0.2455+0.2870$\ir$ & 0 & 2.3015$-$9.5843$\ir$ \\
 0.8901$-$0.2940$\ir$ & 0.5462+0.1296$\ir$ & 0.3637$-$0.1356$\ir$ & $-$1 & 2.3161$-$0.1055$\ir$ \\
 0.1099+0.2940$\ir$ & 0.4538$-$0.1296$\ir$ & 0.6363+0.1356$\ir$ & 1 & 2.3161$-$0.1055$\ir$ \\
 0.4331+0.2763$\ir$ & 0.1239$-$0.2852$\ir$ & 0.7430$-$0.2911$\ir$ & $-$1 & 2.6441$-$4.4197$\ir$ \\
 0.5669$-$0.2763$\ir$ & 0.8761+0.2852$\ir$ & 0.2570+0.2911$\ir$ & 1 & 2.6441$-$4.4197$\ir$ \\
 0.3166$-$0.0393$\ir$ & 0.8325$-$0.2908$\ir$ & 0.9510$-$0.2699$\ir$ & $-$2 & 2.6640$-$5.9503$\ir$ \\
 0.6834+0.0393$\ir$ & 0.1675+0.2908$\ir$ & 0.0490+0.2699$\ir$ & 2 & 2.6640$-$5.9503$\ir$ \\
 0.7711$-$0.3016$\ir$ & 0.5811$-$0.2257$\ir$ & 0.9478+0.2273$\ir$ & $-$1 & 2.7835$-$6.0009$\ir$ \\
 0.2289+0.3016$\ir$ & 0.4189+0.2257$\ir$ & 0.0522$-$0.2273$\ir$ & 1 & 2.7835$-$6.0009$\ir$ \\
 0.9738$-$0.2679$\ir$ & 0.1440+0.3095$\ir$ & 0.3822$-$0.0416$\ir$ & 0 & 2.9472$-$4.5576$\ir$ \\
 0.0262+0.2679$\ir$ & 0.8560$-$0.3095$\ir$ & 0.6178+0.0416$\ir$ & 0 & 2.9472$-$4.5576$\ir$ \\
 0.6743$-$0.2666$\ir$ & 0.7836$-$0.2834$\ir$ & 0.1421$-$0.0500$\ir$ & $-$2 & 3.2342$-$5.9163$\ir$ \\
 0.3257+0.2666$\ir$ & 0.2164+0.2834$\ir$ & 0.8579+0.0500$\ir$ & 2 & 3.2342$-$5.9163$\ir$ \\
 $\frac{\eta}{2}$ & $-\frac{\eta}{2}$ & $\frac{\tau}{2}$ & 1 & 3.3089$-$7.1260$\ir$ \\
 $\frac12$ & $\frac{\tau}{2}$ & $\frac{1}{2}+\frac{\tau}{2}$ & 2 & 4.0896$-$2.5948$\ir$ \\
 $\frac{\tau}{2}$ & 0.5892+0.1294$\ir$ & 0.4108$-$0.1294$\ir$ & 1 & 4.2559$-$2.4972$\ir$ \\
 0.7525$-$0.2958$\ir$ & 0.2475+0.2958$\ir$ & $\frac{1}{2}+\frac{\tau}{2}$ & 1 & 5.3528$-$10.3157$\ir$ \\
 0.3231+0.2849$\ir$ &  0.6769$-$0.2849$\ir$ & $\frac{\tau}{2}$ & 1 & 6.4458$-$12.9612$\ir$ \\
 $\frac12$ & 0.2495+0.2973$\ir$ & 0.7505$-$0.2973$\ir$ & 0 & 6.6317$-$9.7876$\ir$ \\
\hline
\end{tabular}
\caption{Numerical solutions of BAEs (\ref{BAE;M;1})  with $N=6,\,\tau=0.4+0.6\ir,\,\eta=\frac{1}{\eE}+\frac{\ir\pi}{10}$.}\label{4.b}
\end{table}

\begin{hyp}{\label{SingularBR}}
Baxter's $T-Q$ relation (\ref{TQ1}) and the corresponding BAEs (\ref{BAE}) can give the complete spectrum of the Hamiltonian $H$ and the transfer matrix $t(u)$. When $N\geq4$, the BAEs in (\ref{BAE}) have singular physical solutions with two Bethe roots forming a bound pair \cite{Baxter4}
\begin{align}
\l_1=\frac{\eta}{2},\quad \l_2=-\frac{\eta}{2}.
\end{align}
The other Bethe roots $\{\nu_1,\ldots,\nu_{M-2}\}\equiv \{\l_3,\ldots,\l_{M}\}$ satisfy the following BAEs
\begin{align}
&\eE^{2\beta\gamma}\,\left[\frac{\ell{1}(\nu_j+\frac{\eta}{2})}{\ell{1}(\nu_j-\frac{\eta}{2})}\right]^{N-1}\frac{\ell{1}(\nu_j-\frac{3\eta}{2})}{\ell{1}(\nu_j+\frac{3\eta}{2})}\prod_{k\neq j}^{M-2}\frac{\ell{1}(\nu_j-\nu_k-\eta)}{\ell{1}(\nu_j-\nu_k+\eta)}=1,\quad j=1,2,\ldots,M-2.\label{BAE;M;3}\\
&\sin\left(\left(2\sum_{j=1}^{M-2}\nu_j-\beta\tau\right)\pi\right)=0.\label{SelectionRule2}
\end{align} 
\end{hyp}
By substituting the above proposed singular solutions into Eq. (\ref{BAE}), one can readily demonstrate that they satisfy the BAEs (\ref{BAE}).
The $T-Q$ relation corresponding to these singular solutions thus has another deformed form
\begin{align}
\Lambda(u)&=\eE^{\beta\gamma}\frac{\ell{1}^{N-1}(u+\eta)\ell{1}(u-\eta)}{\ell{1}^N(\eta)}\prod_{j=1}^{M-2}\frac{\ell{1}(u-\nu_j-\tfrac{\eta}{2})}{\ell{1}(u-\nu_j+\tfrac{\eta}{2})}+\eE^{-\beta\gamma}\frac{\ell{1}^{N-1}(u)\ell{1}(u+2\eta)}{\ell{1}^N(\eta)}\prod_{j=1}^{M-2}\frac{\ell{1}(u-\nu_j+\tfrac{3\eta}{2})}{\ell{1}(u-\nu_j+\tfrac{\eta}{2})}.
\end{align} 
A bound pair $\l_1=-\l_2=\frac{\eta}{2}$ contributes $-4g(\eta)$ to the energy. With this contribution in mind, the energy can then be expressed as a function of the remaining Bethe roots $\{\nu_1,\ldots,\nu_{M-2}\}$ as follows 
\begin{align}
&E(\nu_1,\ldots,\nu_{M-2})=2\sum_{j=1}^{M-2}[g(\nu_j-\tfrac{\eta}{2})-g(\nu_j+\tfrac{\eta}{2})]+(N-4)g(\eta).\label{Energy2}
\end{align}
\begin{rem}
When we substitute the singular solutions into Eq. (\ref{BetheVector}), the Bethe vector becomes zero. To address this issue, we must rescale the Bethe vector to eliminate the problematic overall factor.   

For $N=2M=4,6,8$, the total number
of the singular solution is $1,4,11$ respectively. 
Such singular solutions also exist in the periodic XXX and XXZ chains \cite{Nepomechie2013,Hao2013}.
\end{rem}

\section{XXZ Limits}\label{XXZ;Limit}

In this section, we will demonstrate how the exact solution of the XXZ chain can be retrieved by taking the limit $\tau\to+\ir\infty$ of the XYZ model.

In the limit $\tau\to+\ir\infty$, the XYZ model degenerates into XXZ model with   \begin{align}
&J_{x,y}\to1,\quad J_z\to\cosh\gamma,\label{XXZlimit}
\end{align} 
and the $R$-matrix becomes the trigonometric one
\begin{align}
R(u)=\begin{pmatrix}
\frac{\sinh(u+\gamma)}{\sinh\gamma} & 0 & 0 & 0 \\
0 & \frac{\sinh u}{\sinh\gamma} & 1 &  0\\
0 & 1 & \frac{\sinh u}{\sinh\gamma} & 0\\
0 & 0 & 0 & \frac{\sinh(u+\gamma)}{\sinh\gamma}
\end{pmatrix}.\label{R2}
\end{align}

The $T-Q$ relation in (\ref{TQ1}) thus  reduces into
\begin{align}
\widetilde{\Lambda}(u)&=\eE^{\beta\gamma}\,\frac{\sinh^N(u+\gamma)}{\sinh^N\gamma}\frac{\widetilde{Q}(u-\gamma)}{\widetilde{Q}(u)}+\eE^{-\beta\gamma}\,\frac{\sinh^Nu}{\sinh^N\gamma}\frac{\widetilde{Q}(u+\gamma)}{\widetilde{Q}(u)},\label{TQ2}\\
\widetilde{Q}(u)&=\prod_{j=1}^{M}\sinh(u-\mu_j+\tfrac{\gamma}{2}).
\end{align}
The corresponding BAEs are
\begin{align}
&\eE^{2\beta\gamma}\,\left[\frac{\sinh(\mu_j+\frac{\gamma}{2})}{\sinh(\mu_j-\frac{\gamma}{2})}\right]^{N}\,\prod_{k\neq j}^{M}\frac{\sinh(\mu_j-\mu_k-\gamma)}{\sinh(\mu_j-\mu_k+\gamma)}=1,\quad j=1,\ldots,M.\label{BAE;XXZ}
\end{align}
One can readily prove that the Bethe roots $\{\l_1,\ldots,\l_M\}$ for the XYZ model and $\{\mu_1,\ldots,\mu_M\}$ for the XXZ model have the following one-to-one correspondence 
\begin{align}
\lim_{\tau\to+\ir\infty}\ir\pi\l_j=\mu_j,\quad j=1,\ldots,M.\label{Correspondence}
\end{align}

To understand how the XYZ model behaves as it approaches the XXZ limit, 
we first analyze the exact solution of XYZ chain in the regime of large ${\rm Im}[\tau]$. For convenience, we let all the Bethe roots lie within the rectangle $$\epsilon\leq {\rm Re}[\l_j]<\epsilon+1,\,\, -\frac{{\rm Im}[\tau]}{2}+\epsilon'\leq {\rm Im}[\l_j]<\frac{{\rm Im}[\tau]}{2}+\epsilon',\quad \epsilon,\epsilon'\in\mathbb{R}$$ where $\epsilon,\epsilon'$ are finite.
From the numerical results illustrated in Tables \ref{Tab2} \ref{Tab3} \ref{Tab4}, we observe that when ${\rm Im}[\tau]$ is large, in some solutions, part of the Bethe roots form a ``quasi-string" as follows
\begin{align}
&\{\l_{l_1},\ldots,\l_{l_m}\}\approx \left\{\pm\frac{\tau}{2}+\kappa,\pm\frac{\tau}{2}+\kappa+\frac{1}{m},\ldots,\pm\frac{\tau}{2}+\kappa+\frac{m-1}{m}\right\},\quad \kappa\in\mathbb{C},\no\\
&2\sum_{k\notin\{l_1,\ldots,l_m\}}\l_j+2m\kappa=m',\quad m'\in\mathbb{Z}.
\end{align} 

Therefore, using Eq. (\ref{Correspondence}), it can be deduced that the Bethe roots $\{\mu_1,\ldots,\mu_M\}$ will divide into two distinct categories: the regular (finite) ones $\{u_1,\ldots,u_{M-m}\}$ and the phantom (infinite) ones \cite{PhantomShort,Zhang2024} $\{v_1,\ldots,v_{m}\}$
or $\{v'_1,\ldots,v'_{m}\}$ where ${\rm Re}[v_j]=+\infty,\,{\rm Re}[v'_j]=-\infty$. The existence of phantom Bethe roots plays a crucial role in the reduction of the XYZ chain to the XXZ chain.

The phantom Bethe roots generate an additional factor to each term in the $T-Q$ relation (\ref{TQ2}). As a consequence, one can rewrite the $T-Q$ relation (\ref{TQ2}) in terms of the finite Bethe roots as
\begin{align}
\widetilde{\Lambda}(u)&=\eE^{(\beta\pm m)\gamma}\,\frac{\sinh^N(u+\gamma)}{\sinh^N\gamma}\prod_{j=1}^{M-m}\frac{\sinh(u-u_j-\frac{\gamma}{2})}{\sinh(u-u_j+\frac{\gamma}{2})}\no\\
&\quad+\eE^{-(\beta\pm m)\gamma}\,\frac{\sinh^Nu}{\sinh^N\gamma}\prod_{j=1}^{M-m}\frac{\sinh(u-u_j+\frac{3\gamma}{2})}{\sinh(u-u_j+\frac{\gamma}{2})},\quad m=0,1\ldots,M,\label{TQ3}
\end{align}
where $+$, $-$ correspond to $\{v_j\}$ and $\{v'_j\}$ respectively. 
\begin{hyp}
The total number of phantom Bethe roots, denoted by $m$, is related to the integer $\beta$ in the $T-Q$ relation (\ref{TQ2}) as $$\beta\pm m=0$$ where $+$, $-$ correspond to $\{v_j\}$ and $\{v'_j\}$ respectively.
\end{hyp}
\begin{proof}
As a polynomial of $u$,  $\widetilde{\Lambda}(u)$ should satisfy the following asymptotic behavior 
\begin{align}
\lim_{u\to +\infty}\widetilde{\Lambda}(u)=\frac{\eE^{Nu}}{(2\sinh\gamma)^N}\left(\eE^{N\gamma-m'\gamma}+\eE^{m'\gamma}\right)+\cdots,\quad m'=0,1,\ldots,N.\label{Asymp;1}
\end{align} 
From Eq. (\ref{TQ3}), we have 
\begin{align}
\lim_{u\to +\infty}\widetilde{\Lambda}(u)=\frac{\eE^{Nu}}{(2\sinh\gamma)^N}\left(\eE^{(\beta\pm m)\eta}\eE^{(N-M+m)\gamma}+\eE^{-(\beta\pm m)\eta}\eE^{(M-m)\gamma}\right)+\cdots.\label{Asymp;2}
\end{align}
From Eqs. (\ref{Asymp;1}) and (\ref{Asymp;2}), one can prove that $\beta\pm m=0$. 
\end{proof}

The aforementioned $T-Q$ relation (\ref{TQ3}) thus is exactly the conventional one given by the algebraic Bethe ansatz method and the number of regular Bethe roots can range from 0 to $\frac{N}{2}$ \cite{Korepin1997}.
The BAEs for the regular Bethe roots $\{u_1,\ldots,u_{M-m}\}$ are
\begin{align}
\left[\frac{\sinh(u_j+\frac{\gamma}{2})}{\sinh(u_j-\frac{\gamma}{2})}\right]^{N}\prod_{k\neq j}^{M-m}\frac{\sinh(u_j-u_k-\gamma)}{\sinh(u_j-u_k+\gamma)}=1,\quad m=0,\ldots,M.\label{BAE;2}
\end{align}  
\begin{hyp}
The phantom Bethe roots $\{v_1,\ldots,v_{m}\}$
or $\{v'_1,\ldots,v'_{m}\}$ will form an equispaced string as follows:
\begin{align}
&v_j=\infty +\frac{\ir\pi j+\ir c}{m},\qquad v'_j=-\infty +\frac{\ir\pi j+\ir c'}{m},\quad j=1,\ldots,m,\quad c,c'\in\mathbb{R}.\label{PhantomString}
\end{align}
\end{hyp}
\begin{proof}
For phantom Bethe roots $\{v_j\}$ and $\{v'_j\}$, the BAEs are satisfied automatically  
\begin{align}
&\eE^{2(\beta+m)\gamma}\prod_{k\neq j}^{m}\frac{\sinh(v_j-v_k-\gamma)}{\sinh(v_j-v_k+\gamma)}=\eE^{2(\beta- m)\gamma}\prod_{k\neq j}^{m}\frac{\sinh(v'_j-v'_k-\gamma)}{\sinh(v'_j-v'_k+\gamma)}=\prod_{k\neq  j}^{m}\frac{\sinh(\ir\pi\,\frac{j-k}{m}-\gamma)}{\sinh(\ir\pi\,\frac{j-k}{m}+\gamma)}\no\\
&=\prod_{k=1}^{m-1}\frac{\sinh(\frac{\ir\pi k}{m}-\gamma)}{\sinh(\frac{\ir\pi k}{m}+\gamma)}=\prod_{k=1}^{m-1}\frac{\sinh(\frac{\ir\pi k}{m}-\gamma)}{\sinh(\ir\pi-\frac{\ir\pi k}{m}-\gamma)}=1.
\end{align}
\end{proof}

The phantom Bethe roots do not contribute to the system's energy, which depends solely on the regular Bethe roots $\{u_1,\ldots,u_{M-m}\}$ 
\begin{align}
E(u_1,\ldots,u_{M-m})=\sum _{j=1}^{M-m} \frac{4 \sinh^2\gamma }{\cosh (2u_j)-\cosh \gamma}+N\cosh \gamma,\quad 
\end{align}

\begin{table}[htbp]
\renewcommand{\thetable}{5}
\begin{minipage}{0.5\textwidth}
\begin{tabular}{|c|c|c|r|}
\hline 
$\l_1$ & $\l_2$ & $\beta$ & $E$~~~~\\
\hline 
 0.0903i & $-$0.0903i & 0 & $-$6.8662 \\
 0 & $-\frac{\tau}{2}$ & $-$1 & $-$4.0393 \\
 0 & $\frac12$+$\frac{\tau}{2}$ & 1 & $-$3.9613 \\
 0 & $\frac12$ & 0 & $-$2.2048 \\
 $-$0.1208i & $\frac12$+0.1208i & 0 & 0.0000 \\
 0.1208i & $\frac12-$0.1208i & 0 & 0.0000 \\
 $-$0.1958i & $-\frac{\tau}{2}+$0.1958i & $-$1 & 0.0000 \\
 0.1958i & $\frac{\tau}{2}-$0.1958i & 1 & 0.0000 \\
 $\frac{1}{2}+\frac{\tau}{2}$$+$0.1873i & $-$0.1873i & 1 & 0.0000 \\
 $\frac{1-\tau}{2}-$0.1873i & 0.1873i & $-$1 & 0.0000 \\
 $\frac{\eta}{2}$& $-\frac{\eta}{2}$ & 0 & 0.0000 \\ 
 0.2475+$\frac{\tau}{2}$ & 0.7525+$\frac{\tau}{2}$ & 2 & 2.2042 \\
$-\frac{\tau}{2}$ & $\frac12-\frac{\tau}{2}$ & $-$2 & 2.2048 \\
$\frac12$ & $-\frac{\tau}{2}$ & $-$1 & 3.9613 \\
$\frac12$ & $\frac12$+$\frac{\tau}{2}$ & 1 & 4.0393 \\
$\frac12-$0.1806i & $\frac12$+0.1806i & 0 & 4.6620 \\
\hline
\end{tabular}
\end{minipage}
\begin{minipage}{0.45\textwidth}
\begin{tabular}{|c|c|c|r|}
\hline 
$\mu_1$ & $\mu_2$ & $\beta$ & $E$~~~~\\
\hline
$-$0.2836 & 0.2836 & 0 & $-$6.8657 \\
0 & $\infty$ & $-$1 & $-$4.0000 \\
0 & $-\infty$+$\frac{\ir \pi }{2}$ & 1 & $-$4.0000 \\
0 & $\frac{\ir \pi }{2}$ & 0 & $-$2.2049 \\
0.3796 & $-$0.3796+$\frac{\ir\pi}{2}$ & 0 & 0.0000 \\
$-$0.3796 & 0.3796+$\frac{\ir\pi}{2}$ & 0 & 0.0000 \\
0.6011 & $\infty$ & $-$1 & 0.0000 \\
$-$0.6011 & $-\infty$ & 1 & 0.0000 \\
$-\infty-\frac{\ir\pi}{2}$ & 0.6011 & 1 & 0.0000 \\
$\infty+\frac{\ir\pi}{2}$ & $-$0.6011 & $-1$ & 0.0000 \\
$\frac{\gamma}{2}$& $-\frac{\gamma}{2}$ & 0 & {0.0000} \\
{$-\infty$+$\frac{\ir \pi }{4}$} & {$-\infty$+$\frac{3\ir \pi }{4}$} & 2 & 2.2049 \\
{$\infty$} & {$\infty$+$\frac{\ir \pi }{2}$} & $-$2 & 2.2049 \\
$\frac{\ir \pi }{2}$ & $\infty$  & $-$1 & 4.0000 \\
$\frac{\ir\pi}{2}$ & $-\infty$+$\frac{\ir \pi }{2}$  & 1 & 4.0000 \\
0.5675+$\frac{\ir\pi}{2}$ & $-$0.5675+$\frac{\ir\pi}{2}$ & 0 & 4.6608 \\
\hline
\end{tabular}
\end{minipage}
\caption{Left: Numerical solutions of BAEs
(\ref{BAE;M;1}) with $N=4,\,\tau=1.8\ir$, $\eta=\frac{\pi}{10}$. Right: Numerical solutions of BAEs (\ref{BAE;XXZ})
with $N=4$, $\gamma=\frac{\ir\pi^2}{10}$.}
\label{Tab3}
\end{table}

\begin{table}[htbp]
\renewcommand{\thetable}{6}
\begin{minipage}{0.5\textwidth}
\begin{tabular}{|c|c|c|c|}
\hline 
$\l_1$ & $\l_2$ & $\beta$ & $E$\\
\hline 
0.9079+0.1098i & 0.0921$-$0.1098i & 0 & $-$6.6424+2.5971i \\
0 & $\frac{1}{2}+\frac{\tau}{2}$ & 1 & $-$4.0385+0.1151i \\
0 & $-\frac{\tau}{2}$ & $-1$ & $-$3.9598$-$0.1151i \\
0 & $\frac{1}{2}$ & 0 & $-$2.4648+4.2288i \\
0.4574+0.1857i & 0.0426$-$0.1857i & 0 & 0.0000 \\
0.5426$-$0.1857i & 0.9574+0.1857i & 0 & 0.0000 \\
$-\frac{\tau}{2}$+0.2990+0.1119i & 0.2010$-$0.1119i & $-$1 & 0.0000 \\
$\frac{\tau}{2}-$0.2990$-$0.1119i & 0.7990+0.1119i & 1 & 0.0000 \\
$-\frac{\tau}{2}$+0.1923$-$0.1081i & 0.8077+0.1081i & $-$1 & 0.0000 \\
$\frac{\tau}{2}$+0.8077+0.1081i & 0.1923$-$0.1081i & 1 & 0.0000 \\
$\frac{\eta}{2}$ & $-\frac{\eta}{2}$ & 0 & 0.0000 \\
$-\frac{\tau}{2}$ & $\frac{1-\tau}{2}$ & $-2$ & 2.4648$-$4.2288i \\
$\frac{\tau}{2}$+0.2447$-$0.0012i & $\frac{\tau}{2}$+0.7553+0.0012i & 2 & 2.4669$-$4.2324i \\
$\frac12$ & $\frac{1}{2}+\frac{\tau}{2}$ & 1 & 3.9598+0.1151i \\
$\frac12$ & $-\frac{\tau}{2}$ & $-1$ & 4.0385$-$0.1151i \\
0.4305$-$0.1932i & 0.5695+0.1932i & 0 & 4.1755+1.6352i \\
\hline
\end{tabular}
\end{minipage}
\begin{minipage}{0.45\textwidth}
\begin{tabular}{|c|c|c|c|}
\hline 
$\mu_1$ & $\mu_2$ & $\beta$ & $E$\\
\hline 
$-$0.3451+2.8523i & 0.3451+0.2893i & 0 & $-$6.6434+2.5956i \\
0 & $-\infty+\frac{\ir\pi}{2}$ & 1 & $-$4.0000 \\
0 & $\infty$ & $-1$ & $-$4.0000 \\
0 & $\frac{\ir\pi}{2}$ & 0 & $-$2.4645+4.2284i \\
$-$0.5833+1.4370i & 0.5833+0.1338i & 0 & 0.0000 \\
0.5833+1.7046i & $-$0.5833+3.0078i & 0 & 0.0000 \\
$\infty$+0.9533i & 0.3457+0.6175i & $-1$ & 0.0000 \\
$-\infty-$0.9533i & $-$0.3457+2.5241i & 1 & 0.0000 \\
$\infty$+0.6175i & $-$0.3457+2.5241i & $-$1 & 0.0000 \\
$-\infty$+2.5241i & 0.3457+0.6175i & 1 & 0.0000 \\
$\frac{\gamma}{2}$ & $-\frac{\gamma}{2}$ & 0 & 0.0000\\
$\infty$& $\infty+\frac{\ir\pi}{2}$ & $-2$ & 2.4645$-$4.2284i \\
$-\infty+\frac{\ir\pi}{4}$& $-\infty+\frac{3\ir\pi}{4}$ & 2 & 2.4645$-$4.2284i \\
$\frac{\ir\pi}{2}$ & $-\infty+\frac{\ir\pi}{2}$ & 1 & 4.0000 \\
$\frac{\ir\pi}{2}$ & $\infty$ & $-1$ & 4.0000 \\
0.6072+1.3527i & $-$0.6072+1.7889i & 0 & 4.1789+1.6327i\\
\hline
\end{tabular}
\end{minipage}
\caption{Left: Numerical solutions of BAEs
(\ref{BAE;M;1}) with $N=4,\,\tau=0.4+1.8\ir$, $\eta=\frac{1}{\eE}+\frac{\ir\pi}{10}$. Right: Numerical solutions of BAEs (\ref{BAE;XXZ}) with $N=4$, $\gamma=\ir\pi(\frac{1}{\eE}+\frac{\ir\pi}{10})$.}
\label{Tab4}
\end{table}  

\begin{rem}
When $m=0$, the regular Bethe roots   $\{u_1,\ldots,u_M\}$ need to satisfy the following sum rule
\begin{align}
&2\sum_{j=1}^{M}u_j=\ir \pi l,\quad\quad l\in\mathbb{Z}.\label{SelectionRule;XXZ}
\end{align}

The presence of the phantom Bethe roots leads to extra degeneracies in the spectrum of the XXZ Hamiltonian. For more examples, see Tables \ref{Tab3} and \ref{Tab4}.

We retrieve the conventional $T-Q$ relation for the periodic XXZ in the limit $\tau\to+\ir \infty$, however the trigonometric limit of the Bethe vector constructed via the generalized algebraic Bethe ansatz method (\ref{BetheVector}) does not coincide with the well known Bethe vector of the XXZ model. 
The inconsistency arises from the degeneracy in the partial spectrum of the transfer matrix $t(u)$, which leads to the non-uniqueness of the corresponding eigenstate.
\end{rem}

\section{Conclusion}
In this paper, we study the generic spin-$\frac12$ XYZ chain under periodic boundary condition. From our numerical data, we conclude that the parameter $\beta$ in the $T-Q$ relation (\ref{TQ1}) and the Bethe vector (\ref{BetheVector}) is an integer, specifically equal to $p$ as indicated in the sum rule (\ref{SelectionRule}). 
Moreover, we demonstrate numerically that Baxter's $T-Q$ relation (\ref{TQ1}) is complete and the corresponding BAEs (\ref{BAE}) have some physical singular solutions, as shown in \ref{SingularBR}.
Additionally, we analyze the XXZ limit of the periodic XYZ chain. In this limit, certain Bethe roots form a phantom string, as described in Eq. (\ref{PhantomString}), and the $T-Q$ relation becomes the conventional one for the XXZ model. Rigorous theoretical analysis are required to substantiate our hypotheses, and this will be the objective in our forthcoming projects.

It is noteworthy that under the following root of unity condition
\begin{align}
P\eta=2K+2L\tau,\quad \quad K,L\in\mathbb{Z},\quad P\in\mathbb{N^+},\label{UnityRoot1}
\end{align} 
the Bethe vector in (\ref{BetheVector}) only contains $P$ terms. The scalar product of the XYZ chain (with even $N$) at the root of unity (\ref{UnityRoot1}), which is independent on the integer $P$, has been thoroughly discussed in Ref. \cite{Slavnov2020}. Since the generic case can be seen as a limiting scenario with $P\to\infty$, we suspect that the scalar product formulas still hold when $\eta$ is not a root of unity, provided that the Bethe roots and the parameter $\beta$ satisfy the BAEs(\ref{BAE}) and the sum rule (\ref{SumRule}).

When there exist another integer $M_0\in[0,N]$ which satisfies 
\begin{align}
&(N-2M_0)\eta=2K_1+2L_1\tau,\quad \quad K_1,L_1\in\mathbb{Z},\quad M_0\neq \tfrac{N}{2},\label{UnityRoot2}
\end{align}
a homogeneous $T-Q$ relation with $M_0$ Bethe roots exists and the generalized algebraic Bethe ansatz method remains applicable, regardless of whether the total site number is odd or even. Under the condition (\ref{UnityRoot2}), the hypotheses presented in this paper might not hold validity. A straightforward argument is that the sum rule shown in (\ref{SelectionRule}) does not exist when $M_0\neq\frac{N}{2}$ \cite{Zhang2024}.

For the generic periodic XYZ chain with odd site number, the exact spectrum of the transfer matrix $t(u)$ can be given by an inhomogenous $T-Q$ relation \cite{Wang-book,Cao2014,Xin2020}. However, constructing the corresponding Bethe vector remains an unresolved open question within the field.

Another interesting issue is proposing a fast numerical method to solve the elliptic type BAEs (\ref{BAE}). The correspondence between the XYZ model and the XXZ model, as discussed in Section \ref{XXZ;Limit}, indeed makes it more practical to solve the BAEs once we have the exact solution of the periodic XXZ model.

\section*{Acknowledgments}
The author acknowledges the financial support from the National Natural Science Foundation of China (No. 12204519).

\end{document}